\documentclass[conference]{IEEEtran}
\usepackage{amssymb}
\usepackage{amsmath}
\usepackage{epsfig}
\usepackage{rotating}

\begin{document}
\newtheorem{theorem}{Theorem}
\newtheorem{corollary}{Corollary}
\newtheorem{definition}{Definition}
\newtheorem{property}{Property}
\newtheorem{lemma}{Lemma}

\newcommand{\define}{\stackrel{\triangle}{=}}

\pagestyle{empty}

\def\QED{\mbox{\rule[0pt]{1.5ex}{1.5ex}}}
\def\proof{\noindent\hspace{2em}{\it Proof: }}

\date{}
\title{Sum-Capacity and the Unique Separability of the Parallel Gaussian MAC-Z-BC Network} 
\author{\authorblockN{Viveck R. Cadambe, Syed A. Jafar}
\authorblockA{Electrical Engineering and Computer Science\\
University of California Irvine, \\
Irvine, California, 92697, USA\\
Email: {vcadambe@uci.edu, syed@uci.edu}\\ }}
\maketitle

\def\QED{\mbox{\rule[0pt]{1.5ex}{1.5ex}}}
\def\proof{\noindent\hspace{2em}{\it Proof: }}

\newcommand{\xH}{H}
\newcommand{\xX}{X}
\newcommand{\xY}{Y}
\newcommand{\xZ}{Z}

\newcommand{\lSigma}{\rotatebox[origin=c]{180}{$\Sigma$}}

\newcommand{\uH}{\underline{H}}
\newcommand{\uX}{\underline{X}}
\newcommand{\uY}{\underline{Y}}
\newcommand{\uZ}{\underline{Z}}
\renewcommand{\baselinestretch}{0.95}

\begin{abstract} 
It is known that the capacity of parallel (e.g., multi-carrier) Gaussian point-to-point, multiple access and broadcast channels can be achieved by separate encoding for each subchannel (carrier) subject to a power allocation across carriers. Recent results have shown that parallel interference channels are not separable, i.e.,  joint coding is needed to achieve capacity in general. This work studies the separability, from a sum-capacity perspective, of single hop Gaussian interference networks with independent messages and arbitrary number of transmitters and receivers. The main result is that the only network that is always (for all values of channel coefficients) separable from a sum-capacity perspective is the MAC-Z-BC network, i.e., a network where a MAC component and a BC component are linked by a Z component. The sum capacity of this network is explicitly characterized.
\end{abstract}

\section{Introduction}

\allowdisplaybreaks{
Wireless networks are often associated with channel states that are time-varying/frequency-selective. These networks are equivalently described as parallel (i.e. multi-carrier) Gaussian networks. 
The \emph{separability} of a parallel Gaussian network implies that the network capacity can be achieved by using separate coding/decoding over each carrier (i.e. parallel component), so that the capacity can be expressed as a sum of the capacities of the individual sub-channels, subject to optimum power allocation. For instance, it is well known that parallel Gaussian point to point channels are separable, with the optimal power allocation across the carriers found using the water-filling algorithm. Similarly, the parallel multiple access (MAC) and broadcast (BC) networks are known to be separable. An important consequence of this separability is that it greatly simplifies the problem of its capacity characterization of parallel channels because for a separable network it suffices to study the network in a fixed channel state. Specifically, the capacities of ergodic fading Gaussian point to point, multiple access and broadcast networks are the averages of the capacities achieved over each channel fading state, subject to the optimum  power allocation.  
\begin{figure}
\begin{center}
\epsfig{file=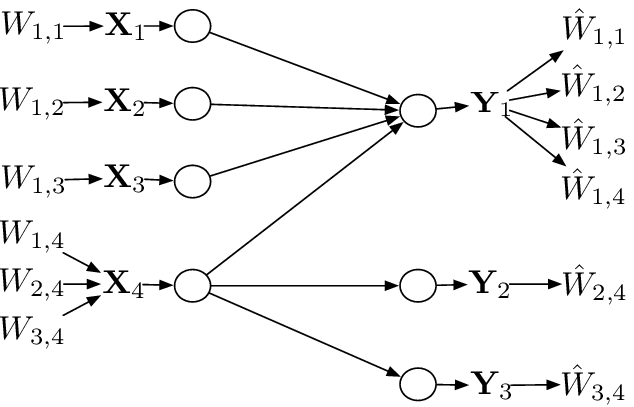}
\caption{A MAC-Z-BC network with $S=4$ transmitters and $D=3$ receivers}
\label{fig:maczbc}
\end{center}
\vspace{-25pt}
\end{figure}

Unlike parallel Gaussian point to point, MAC and BC networks, it has been discovered recently, in references \cite{Cadambe_Jafar_inseparable, Sankar_inseparable} that parallel Gaussian interference networks are not separable, i.e., parallel interference channels in general need \emph{joint} coding over the carriers to achieve capacity. Thus, for these networks the capacity characterization for a fixed channel state does not provide a direct extension to the capacity of parallel, e.g., time-varying/frequency-selective scenarios. In particular, the ergodic capacity of interference networks can be much higher than the average of the capacity expressions for each fixed channel state, even with the best power allocation across states \cite{Sankar_inseparable,Nazer_Gastpar_Jafar_Vishwanath}. 



Beyond the special cases identified above, i.e., the Gaussian MAC, BC, and interference networks, for the vast majority of Gaussian networks the separability properties have not been determined. As the focus of research in network information theory advances through increasingly complex network topologies, the separability properties of these networks will be the key to their capacity characterizations in the presence of fading. It is this endeavor - the study of separability properties of more complex (albeit single-hop) Gaussian wireless networks - that is the focus of this paper. Our main result can be summarized as follows. \emph{The only distributed\footnote{We assume that nodes are half-duplex, i.e., transmitters cannot receive and receivers cannot transmit, so that co-operation, feedback, relaying are all precluded in the network.} single-hop wireless network, which is separable from a sum-capacity perspective, is the MAC-Z-BC network (Fig. \ref{fig:maczbc}, defined formally later).} The separability of point-to-point MAC and BC networks automatically follow from our main result, since they are special cases of the MAC-Z-BC network. The separability leads to a characterization of the sum-capacity of the parallel MAC-Z-BC network. 

\section{System Model : Parallel Single-hop Wireless Networks}
\begin{definition}{ \bf Network} : A Gaussian wireless single-hop network $\mathcal{N}$ is characterized by $({S},{D},\mathcal{E},\mathcal{M})$. Here $S$ denotes the number of sources or transmitters, $D$ represents the number of destinations or receivers. The topology of the network is identified by a \emph{connected} undirected bipartite graph between the set of transmitters $\mathcal{S}=\{1,2,\ldots,S\}$ and the set of receivers $\mathcal{D}=\{1,2,\ldots,D\}$, with $\mathcal{E} \subseteq \mathcal{D} \times \mathcal{S}$ representing the set of edges in the network. $\mathcal{M}\subseteq \mathcal{E}$ is used to identify the \emph{message-graph or message-set}, where corresponding to element $(j,i) \in \mathcal{M}$, $j$ is termed the message destination, and $i,$ the message source.  Finally, we assume that $1 \leq i \leq S \Rightarrow \exists j_0, (j_0,i) \in \mathcal{M}$ and $1 \leq j \leq D \Rightarrow \exists i_0, (j,i_0) \in \mathcal{M}.$ In other words, for every source, there exists at least one message destination, and similarly, for every destination, there exists at least one message source.
\end{definition}

Given a network $\mathcal{N}\define ({S},{D},\mathcal{E},\mathcal{M})$, an \emph{instance} of the network is uniquely identified by $(F,\overline{P},\overline{\mathbf{H}})$, where $F$ denotes the number of carriers, $\overline{P}=(P_1,P_2,\ldots,P_S)$ is a $S \times 1$ vector denoting the power constraints and $\mathbf{\overline{H}}$ is a $DF\times SF$ dimensional complex channel gain matrix. The channel gain matrix can be expressed as a block matrix, $\mathbf{\overline{H}}=\left[ \mathbf{H}_{j,i}\right]$ where $\mathbf{H}_{j,i}$ is a diagonal\footnote{Note that the diagonal nature of the channel matrix reflects the fact that we assume negligible inter-carrier interference.} $F \times F$ dimensional matrix for $(j,i) \in \{1,2,\ldots,D\}\times\{1,2,\ldots,S\}$, such that $\mathbf{H}_{j,i}=\mathbf{0}$ if $(j,i) \notin \mathcal{E}$.
We use the terms \emph{instance of the network}, and \emph{channel}, interchangeably in the paper.
  The input-output relations of the channel can be represented as 
\begin{equation}
\mathbf{Y}_j(\tau) = \sum_{(j,i) \in \mathcal{E}} \mathbf{H}_{j,i} \mathbf{X}_i(\tau) + \mathbf{Z}_j(\tau)
\end{equation}
where, corresponding to the $\tau$th channel use, $\mathbf{X}_i(\tau)=\left(X_i(1,\tau),X_i(2,\tau),\ldots,X_i(F,\tau)\right)$ represents the $F \times 1$ complex input vector at Transmitter $i \in \{1,2,\ldots,S\}$, whose $f$th component indicates the input corresponding to the $f$th carrier for $f \in \{1,2,\ldots,F\}.$ Similarly, for $j \in \{1,2,\ldots,D\}$, $\mathbf{Y}_j(\tau)$ and $\mathbf{Z}_j(\tau)$ are $F \times 1$ vectors respectively representing the complex output and circularly symmetric complex additive white Gaussian noise respectively. We assume that the additive noise has zero-mean and identity covariance. For brevity of notation, the dependence on the channel-use index $\tau$ is dropped unless necessary. The $f$th diagonal entry of diagonal matrix $\mathbf{H}_{j,i}$ represented by $H_{j,i}(f)$ indicates the channel gain between Transmitter $i$ and Receiver $j$ in the $f$th carrier. There are $|\mathcal{M}|$ independent messages in the system indicated by $W_{j,i}$ where $(j,i) \in \mathcal{M}$. Message $W_{j,i}$ is generated at the Transmitter $i$ and is intended for Receiver $j$. We assume that the codewords satisfy an average power constraint, i.e., $ E \left[ \frac{1}{T} \sum_{t=1}^{T} \sum_{f=1}^{F}  |X_{j}(f,t)|^2\right] \leq P_j,$ where $T$ denotes the length of the codeword. For codewords of length $T$, the rates $R_{j,i}=\frac{\log\left(|W_{j,i}|\right)}{T}$ are said to be \emph{achievable} if the probability of error of all messages can be made arbitrarily small by choosing an appropriately large $T$. The capacity region of the channel $\mathcal{C}^{\mathcal{N}}(F,\overline{P}, \overline{\mathbf{H}})$ is defined to be the set of all achievable rate matrices $\left[R_{j,i}\right]$ over the channel. The sum-capacity of the channel $C^{\mathcal{N}}(F,\overline{P},\overline{\mathbf{H}})$ is defined as 
$$ C^{\mathcal{N}}(F,\overline{P},\overline{\mathbf{H}})= \max_{[R_{j,i}] \in \mathcal{C}^{\mathcal{N}}(F,\overline{P},\overline{\mathbf{H}})} \sum_{(j,i) \in \mathcal{M}} R_{j,i}.$$
Note that the above framework captures any arbitrary parallel single-hop wireless network with distributed single-antenna half-duplex transmitting and receiving nodes. For example, the $K$-user interference network is characterized by $(S=K,D=K,\mathcal{E},\mathcal{M})$, where $\mathcal{E}=\{1,2,\ldots,K\}^2$ and $\mathcal{M}=\{(i,i):i \in \{1,2,\ldots,K\}\}$. Following standard nomenclature, we refer to the edges $(i,i)$ as the direct links, and the edges $(i,j), i \neq j$ as cross links for $i,j \in \{1,2,\ldots,K\}$.
\subsubsection*{Notation}
The notation $\overline{H}(f)$ is used to indicate the $D \times S$ matrix of channel gains corresponding to the $f$th carrier, i.e., $\left[ H_{j,i}(f) \right]$, where $j=1,2,\ldots,D, i=1,2,\ldots,S$.  When $F=1$, we use $\overline{H}$ to represent the matrix $\overline{H}(1)$. We also use the notation $\mathbf{X}_i^{T}=(\mathbf{X}_i(1),\mathbf{X}_i(2),\ldots,\mathbf{X}_i(T))$ for $1 \leq i \leq S$. The quantities $\mathbf{Y}_j^{T}, \mathbf{Z}_j^{T},X_i^T(f), Y_j^{T}(f),Z_j^{T}(f)$ are defined similarly for $1 \leq i \leq S, 1 \leq j \leq D, 1 \leq f \leq F$.

\subsection{Separability}
\begin{definition}{\bf Separability of a Network}
A channel belonging to network $\mathcal{N}$ characterized by $(F,\overline{P},\overline{\mathbf{H}})$ is said to be \emph{separable} if and only if
$$ C^{\mathcal{N}}(F,\overline{P},\overline{\mathbf{H}}) = \max_{\sum_{f=1}^{F}\overline{P}^{[f]} \leq \overline{P} }\sum_{f=1}^{F}C^{\mathcal{N}}\left(1,\overline{P}^{[f]},\overline{H}^{[f]}\right)$$
A channel is said to be \emph{inseparable}, if it is not separable. A network $\mathcal{N}$ is said to be {separable} if all its instances (channels) are separable. The network is {inseparable} if it is not separable.
\end{definition}

\subsubsection*{Remark}
We focus on separability from the perspective of \emph{sum}-capacity of the network in this paper. Separability of networks from the perspective of the whole capacity region can be correspondingly defined; their study is an area of future work.

\begin{definition} {\bf Sub-network} A network $\mathcal{N}^{'} \define ({S}^{'},{D}^{'},\mathcal{E}^{'},\mathcal{M}^{'})$ is said to be a \emph{sub-network} of network $ \mathcal{N}\define({S},{D},\mathcal{E},\mathcal{M})$ if ${S}^{'} \leq {S}, {D}^{'} \leq {D}, \mathcal{E}^{'} \subseteq \mathcal{E}$ and $\mathcal{M}^{'}=\mathcal{M} \cap \mathcal{E}^{'}$.
\end{definition}
In plain words, sub-network $\mathcal{N}^{'}$ can be obtained from network $\mathcal{N}$ by deleting certain links from network $\mathcal{N}$.  For example, the $Z$-interference network, characterized by $(S=2,D=2,\mathcal{E}=\{(1,1),(1,2),(2,2)\}, \mathcal{M}=\{(1,1),(2,2)\}$ is a sub-network of $K$-user interference networks.

\begin{lemma}
\label{lem:subnet}
A network is separable only if all its sub-networks are separable. Equivalently, if a sub-network $\mathcal{N}^{'}$ of network $\mathcal{N}$ is inseparable, then the network $\mathcal{N}$ is also inseparable.
\end{lemma}
The above lemma can be shown based on the definition of separability as follows. First, note that $\mathcal{N}^{'}$ can be derived by setting certain channel gains in $\mathcal{N}$ to zero. Consider the case where $\mathcal{N}^{'}$ is inseparable. The inseparability of $\mathcal{N}^{'}$ implies that an instance of $\mathcal{N}^{'}$ exists where separate coding is sub-optimal. This instance can be used to construct an instance of $\mathcal{N}$ where separate coding is suboptimal, by setting the appropriate channel gain matrices to $\mathbf{0}$ in $\mathcal{N}$. Hence $\mathcal{N}$ is inseparable.



\section{Main Result : Unique separability of the MAC-Z-BC channel}
\subsection{Background on Interference Networks and Motivation}
\label{sec:background}
Interference networks have been shown to be inseparable in \cite{Cadambe_Jafar_inseparable, Sankar_inseparable}. We summarize the main factors behind the inseparability of interference networks below, since they will be seen to recur in our main result as well. 
\begin{itemize}
\item \cite{Cadambe_Jafar_inseparable} Joint coding enables interference alignment, whereas separate coding does not (used to show inseparability of the $3$-user interference network).
\item \cite{Sankar_inseparable} Joint coding enables a receiver to use the interfering signal received over certain carriers to cancel interference from other carriers - this is not possible with separate coding (used to show that the $Z$-interference network is inseparable).
\end{itemize}

Note that the interference network does not exploit the full potential of the physical channel because every link does not carry a message. For example, the capacity of the $Z$-interference channel does not increase (beyond an extent) as the strength of the cross-link is made arbitrarily large, if the strength of the direct links are held constant. Based on this motivation, we study the separability properties of a more general class of networks where every link can carry a message, i.e., where $\mathcal{M}$ can be any sub-set of $\mathcal{E}.$ We will find that, surprisingly, the $Z$ network - which is physically identical to the $Z$-interference network except that the cross-link also carries a message ($\mathcal{M}=\mathcal{E}$) - is in fact separable from a sum-capacity perspective. 

\subsection{Main Result : Unique Separability and Sum-Capacity of MAC-Z-BC network}

\begin{definition}
The MAC-Z-BC network is characterized by $(S,D,\mathcal{E},\mathcal{M}=\mathcal{E})$, where $\mathcal{E}$ has the following property.
$$\deg(T_i) > 1, i^{'} \neq i \Rightarrow \deg(T_{i^{'}}) = 1, \forall i^{'},i \in \mathcal{S}$$
$$\deg(R_j) > 1, j^{'} \neq j \Rightarrow \deg(R_{i^{'}}) = 1, \forall j,j^{'} \in \mathcal{D} $$
where, $\mathcal{S}=\{1,2,\ldots,S\}$, $\mathcal{D}=\{1,2,\ldots,D\}$, $\deg(T_i)$ is the degree of Transmitter $i$, and $\deg(R_j)$ is the degree of Receiver $j$.
\end{definition}

\subsubsection*{Remark}
The MAC-Z-BC network, based on the above definition, can be noted to contain a MAC component and a BC component connected by a $Z$ component (See Fig. \ref{fig:maczbc}). The MAC, BC and $Z$ networks are sub-networks of the MAC-Z-BC network.

The main result of the paper can be stated as follows : 
\begin{theorem}
\label{thm:main}
A network $\mathcal{N}$ is separable if and only if it is the MAC-Z-BC network (or one of its sub-networks).
\end{theorem}
\begin{theorem}
\label{thm:single-carrier}
Consider a single-carrier MAC-Z-BC channel characterized by $(1, \overline{P},\overline{H})$, where $\deg(T_i) > 1 < \deg(R_j) \Rightarrow j=1, i=S$ (See Fig. \ref{fig:maczbc}).  Then, its sum-capacity is
\begin{eqnarray*} C^{MAC-Z-BC}(1,\overline{P},\overline{H}) &=& \log\left( 1+ \sum_{j=1}^{S} |H_{1,j}|^2 P_j\right)\\&&+ \log\left(\frac{1+|H|^2 P_S}{1+|H_{1,S}|^2 P_S}\right),\end{eqnarray*}
where $H=\max_{j=1,2,3,\ldots,D}|H_{j,S}|$.
\end{theorem}
The two theorems lead to a sum-capacity characterization of the parallel MAC-Z-BC network as the sum of the sum-capacities of the individual carriers, under an optimal power allocation. The achievability of the expression in Theorem \ref{thm:single-carrier} can be seen as follows. Let $j^{*}=\arg\max_{j=1,2,\ldots,D} H_{j,S}$ so that $|H|=H_{j^{*},S}$. Transmitter $S$ generates only one message $W_{j^{*},S}$ and sets all other messages to null, i.e., $W_{j,S}=\phi, j \neq j^{*}$. If $j^{*}=1$, then the capacity can be achieved over the multiple access channel formed at Receiver $1$ and Transmitters $1,2,\ldots,S$. If $j^{*}\neq 1$, Receiver $j^{*}$ observes a point-to-point channel from Transmitter $S$ to achieve a rate of $\log(1+|H_{j^{*},S}|^2 P_S)$. Receiver $1$ treats the interference from Transmitter $S$ as noise and a total rate of $\log(1+\sum_{i=1}^{S-1}|H_{1,i}|^2 P_i)$ can be achieved for the messages $W_{1,j},j\neq S$ over the multiple access channel formed at the receiver. The converse for Theorem \ref{thm:single-carrier} follows from the converse of Theorem \ref{thm:main} (Appendix \ref{app:separable}). 

The proof of Theorem \ref{thm:main} involves two parts. First, we prove that the MAC-Z-BC network is separable; then, we show that any network which is not the MAC-Z-BC network is inseparable. The proof of the separability of the MAC-Z-BC network is placed in Appendix \ref{app:separable}. Intuitively, the separability of the $Z$ network (which is a sub-network of the MAC-Z-BC network) can be understood as follows. Note that the $Z$-interference network is inseparable because joint coding enables better interference-cancellation over the cross-link. However, in the $Z$ network, any bit that the receiver is able to decode over the cross-link can be used for the desired message, rather than for interference management and the second factor of inseparability listed in Section \ref{sec:background}, is avoided from a sum-capacity perspective. We will now present a proof of the second part of Theorem \ref{thm:main}, i.e., we show that any network which is not the MAC-Z-BC network is inseparable.
 
\begin{property} Consider any network $\mathcal{N}$ characterized by $(S,D,\mathcal{E},\mathcal{M}=\mathcal{E})$. In the network, if $\exists (i,j), i \neq j$ such that 
$$\deg(T_i) > 1 < \deg(T_j)\mbox{ or } \deg(R_i) > 1 < \deg(R_j)$$
then, the network $\mathcal{N}$ has at least one of the following three networks as a sub-network.
\begin{enumerate}
\item The $Z$-interference network.
\item The $2$-user $X$ network (Fig. \ref{fig:Xinsep}(a)), with $S=2,D=2,\mathcal{E}=\{1,2\}\times\{1,2\}$, $\mathcal{M}=\mathcal{E}$.
\item The $\Sigma$ network (Fig. \ref{fig:Xinsep}(b)), with $S=2,D=3,\mathcal{E}=\{(1,1),(2,1),(2,2),(3,2)\}$, $\mathcal{M}=\mathcal{E}$.
\item The $\lSigma$ network (Fig. \ref{fig:Xinsep}(c)), with $S=3,D=2,\mathcal{E}=\{(1,1),(1,2),(2,2),(2,3)\}$, $\mathcal{M}=\mathcal{E}$.
\end{enumerate}
\end{property}

\begin{figure*}
\begin{center}
\epsfig{file=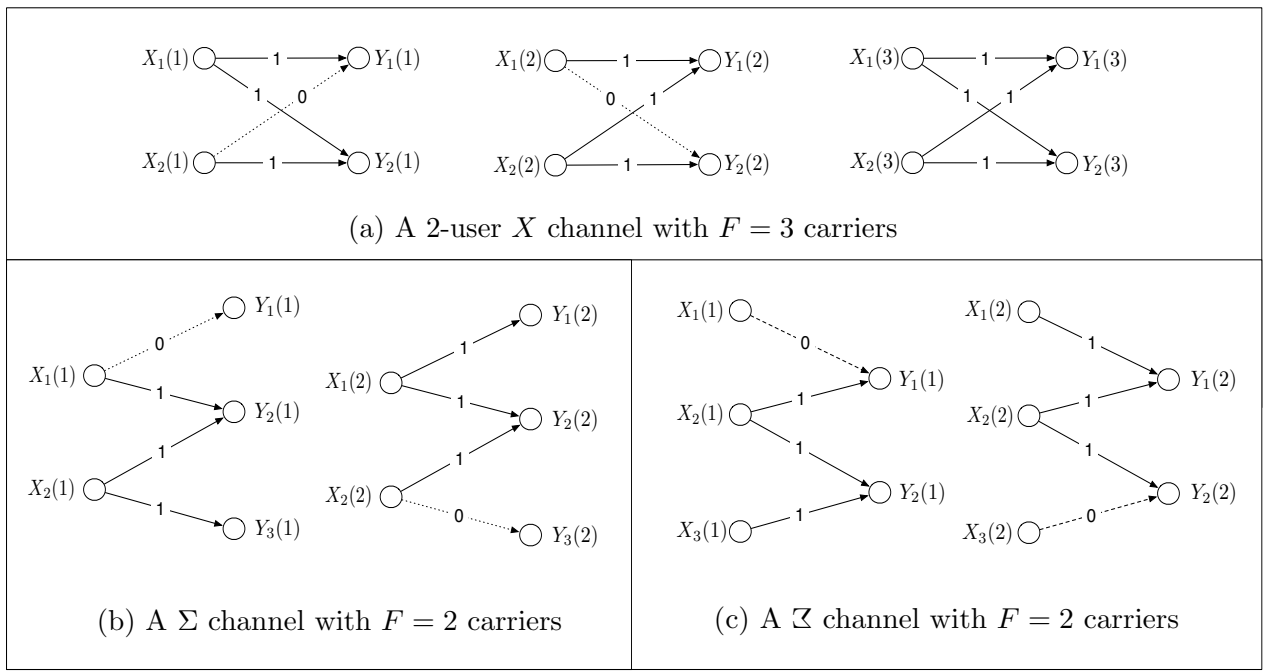,height=3.5in, width=6.6 in}
\caption{Inseparable wireless channels (Note that the number on the link denotes the channel gain).}
\label{fig:Xinsep}
\end{center}
\vspace{-26pt}
\end{figure*}

Based on the above property of connected bipartite graphs (See Appendex \ref{app:property} for a proof) and Lemma \ref{lem:subnet}, it is enough to show that the $Z$-interference network, $2$-user $X$ network, the $\Sigma$ network and the $\lSigma$ networks are all inseparable. The $Z$-interference network has been shown to be inseparable in \cite{Sankar_inseparable}. We show the inseparability of the latter networks below. 

\subsubsection{The $2$-user $X$ network}
Consider a $2$-user $X$ channel (See Fig. \ref{fig:Xinsep} (a)) where $F=3$, $\overline{P}=(P,P)$ and 

$$\overline{H}(1)=\left[\begin{array}{cc}1& 0 \\ 1& 1\end{array} \right]=(\overline{H}(2))^T, \overline{H}(3)=\left[\begin{array}{cc}1 &1 \\ 1& 1\end{array} \right]$$
With this, it can be noted (because of the fact that the first two carriers form $Z$ channels, and the third carrier is a degraded $X$ channel \cite{Cadambe_Jafar_maczbc}) that
\begin{equation*}  C^{X}(1,\overline{P},\overline{H}(f))\leq \log(P)+o(\log(P)) \end{equation*}
\begin{align}
 &\Rightarrow \max_{\sum_{f=1}^{3}\overline{P}^{[f]}\leq \overline{P}}\sum_{f=1}^{3} C^{X}(1,\overline{P}^{[f]},\overline{H}(f))\nonumber\\& \leq  3 \log(P)+o(\log(P)),\label{eq:xouterbound}
\end{align}
Joint coding can achieve a higher rate (scaling with $P$). To see this
    consider an achievable scheme, where each message $W_{i,j}, i,j \in \mathcal{M}$ is encoded as a Gaussian codeword $x_{i,j}(\tau), \tau=1,2,\ldots,T$. The transmit codeword at Transmitter $j$, $\mathbf{X}_j,$ is determined as 
\begin{equation} \mathbf{X}_i(\tau) = \sum_{j=1}^{D} x_{j,i}(\tau) \mathbf{V}_{j,i}\label{eq:beamforming}\end{equation}
where $\tau=1,2,\ldots,T$, and $\mathbf{V}_{j,i}$ is a $F \times 1$ (beamforming) vector as follows.
$$ \mathbf{V}_{1,1} = \mathbf{V}_{1,2}=\left[1~~0~~1\right], \mathbf{V}_{2,2} = \mathbf{V}_{2,1}=\left[0~~1~~1\right]$$
The above beamforming vectors ensure that $\mathbf{H}_{1,1} \mathbf{V}_{2,1}=\mathbf{H}_{1,2} \mathbf{V}_{2,2}$ and $\mathbf{H}_{2,1} \mathbf{V}_{1,1}=\mathbf{H}_{2,2} \mathbf{V}_{1,2}$, i.e., both interfering messages at each receiver are \emph{aligned} into one dimension, similar to \cite{Jafar_Shamai}. It can also be verified that, at Receiver $j\in\{1,2\}$, the two desired signal vectors  are linearly independent of the (aligned) interference. Thus in a $F=3$ dimensional space, by linearly nulling the (aligned) interference dimension, one interference-free dimension can be achieved for each of the two desired messages at a receiver. Each message can thus achieve a rate of $\log(P)+o(\log(P))$ \cite{Jafar_Shamai} and
\begin{equation*}  C^{X}(1,\overline{P},\overline{\mathbf{H}})\geq 4\log(P)+o(\log(P)). \end{equation*}
Comparing the above with (\ref{eq:xouterbound}), we can conclude that, for sufficiently large $P$, the $2$-user $X$ channel is inseparable. Thus, the enabling of alignment via joint coding renders the $2$-user $X$ network inseparable.
\subsubsection{The $\Sigma$ network}

The inseparability of the $\Sigma$ channel where $F=2, \overline{P}=(P,P)$ and the channel gains shown in Figure \ref{fig:Xinsep} (b) can be shown based on the enabling of interference alignment. To see this, first, note that separate coding allows, at most, a sum-rate of $2 \log(P)+o(\log(P))$, since each carrier viewed separately forms a $Z$ channel whose sum-rate is at most $\log(P)+o(\log(P))$ per carrier \cite{Chong_Motani_Garg_Z}.
A sum-rate of $3\log(P)+o(\log(P))$ based on a joint coding scheme, with each of $W_{1,1}, W_{2,1},W_{3,2}$ achieving a rate of $\log(P)+o(\log(P))$ (and $R_{1,2}=0)$). The messages are encoded as in (\ref{eq:beamforming}), where $\mathbf{V}_{1,1}=\mathbf{V}_{2,3}=[1~~0]^T$ and $\mathbf{V}_{2,1}=[1~~1]^T$. Then, the two interfering messages $W_{1,1},W_{3,2}$ align at Receiver $2$, i.e. $\mathbf{H}_{2,1}\mathbf{V}_{1,1}=\mathbf{H}_{2,2}\mathbf{V}_{3,2}$. It can be verified that this scheme achieves one interference-free dimension for $W_{1,1},W_{2,1},W_{3,2}$ so that the desired rate (scaling with $P$) is achieved implying the inseparability of the $\Sigma$ network. 
\\\indent \emph{3) The $\lSigma$ network:}
The $\lSigma$ channel of Fig. \ref{fig:Xinsep}(c), which is a reciprocal of the $\Sigma$ channel of Fig. \ref{fig:Xinsep} (b) can be shown to be inseparable using the reciprocity of beamforming based alignment schemes (\cite{Cadambe_Jafar_X, Cadambe_Jafar_maczbc}). Note that in the $\lSigma$ network, each receiver faces at most one interferer, and it is hard to conceive of the possibility of alignment. We take a closer look at the channel to understand its inseparability from the perspective of the factors listed in Section \ref{sec:background}. First note that separate coding achieves a sum-rate of at most $2 \log(P)+o(\log(P))$, since each carrier forms a $Z$ channel. A sum-rate of $3\log(P)+o(\log(P))$ can be achieved using joint coding with each of $W_{1,1},W_{1,2},W_{2,3}$ achieving $1$ interference-free dimension. To see this, let Transmitter $2$ encode $W_{1,2}$ so that it transmits identical signals on both carriers - this would be a beamforming based joint coding scheme where the beamforming vector is $[1~~1]^T$. Then, each of the two receivers can subtract the signals received along both carriers to null the signal from Transmitter $2$ to achieve an interference-free dimension for $W_{1,1},W_{2,3}$. Note that Receiver $1$ gets an interference-free dimension for $W_{1,2}$ on the first carrier. Thus, joint coding enables better interference cancellation at the two receivers and the second factor identified in Section \ref{sec:background} causes inseparability. While the fundamental reason of inseparability remains the same as the $Z$-interference channel \cite{Sankar_inseparable}, the nature of interference-cancellation in our example differs from the \cite{Sankar_inseparable} in that it is linear, whereas the scheme in \cite{Sankar_inseparable} is non-linear (successive decoding).  Note that unlike the $Z$ network where interference management (and the associated factor of inseparability) can be avoided by using the cross-link for the desired message, interference management is unavoidable in the $\lSigma$ network because $W_{1,2}$ and $W_{2,1}$ are necessarily an interfering messages respectively for Receivers $2$ and $1$.


\section{Conclusion}
We recognize two principle factors causing inseparability in parallel wireless networks and based on the intuition obtained, identify the MAC-Z-BC as the uniquely separable wireless network. Associated interesting areas of future work are identification of separability properties of networks with multiple antennas, and its study from the perspective of the entire capacity region.
\appendices 
\section{The MAC-Z-BC channel is separable}
\label{app:separable}
We assume that $\deg(T_i) > 1 < \deg(R_j) \Rightarrow j=1, i=D$ (See Fig. \ref{fig:maczbc}). We show a (tight) converse, which bounds the sum-rate on the parallel MAC-Z-BC channel. The achievability of the sum-rate upper bound follows from separate coding and the capacity of the single-carrier MAC-Z-BC (Theorem \ref{thm:single-carrier}). 
For the converse, we create Receiver $D+1$ whose output is 
$$ \mathbf{Y}_{D+1}(\tau)= \mathbf{H}\mathbf{X}_S(\tau)+\mathbf{Z}_{D+1}(\tau)$$
where $\mathbf{Z}_{D+1}$ is unit-variance circularly symmetric white Gaussian noise. $\mathbf{H}$ is a $F \times F$ diagonal matrix whose $f$th diagonal entry is determined as $H(f)=\max_{i=1,\ldots,D} H_{i,S}(f)$. Note that Receiver $D+1$ is an enhanced receiver so that, with any achievable scheme, it can decode $W_{i,S}, \forall i=1,2,\ldots,D$. Given any achievable scheme of length $T$, we denote $P_j^{[f]}=E\left[\frac{1}{T} |X_j(f)|^2\right],j=1,2,\ldots,S$. Note that $\sum_{f=1}^{F} P_j^{[f]} < P_j$. Using Fano's inequality for we can write for any $\epsilon > 0$
\begin{eqnarray}
\sum_{i=1}^{D} R_{i,S} - T\epsilon &\leq& I(\mathbf{Y}_{D+1}^T;\mathbf{X}_{S}^T)\nonumber\\&=& h(\mathbf{H} \mathbf{X}_S^T+\mathbf{Z}_{D+1}^T)-h(\mathbf{Z}_{D+1}^T)\label{eq:1}
\end{eqnarray}
\begin{eqnarray}
\sum_{j=1}^{S-1} R_{1,j} - T\epsilon &\leq& I\left(\mathbf{Y}_1^{T};\mathbf{X}_1^T,\mathbf{X}_2^T,\ldots,\mathbf{X}_{S-1}^{T}\right) \nonumber \\
&=& h(\mathbf{Y}_1^T)-h(\mathbf{H}_{1,S}\mathbf{X}_{S}^T+\mathbf{Z}_{1}^{T}) \label{eq:2}
\end{eqnarray}
Adding (\ref{eq:1}),(\ref{eq:2}), we get
\begin{align*}
&\sum_{(i,j)\in \mathcal{M}} R_{i,j}-2T\epsilon \nonumber \\ 
&\leq h(\mathbf{Y}_1^T)-h(\mathbf{Z}_{D+1}^T)\nonumber \\&+ h(\mathbf{H} \mathbf{X}_S^T+\mathbf{Z}_{D+1}^T) - h(\mathbf{H}_{1,S}\mathbf{X}_{S}^T+\mathbf{Z}_{1}^{T}) \nonumber\\
&\leq T\sum_{f=1}^{F} \Big(h({Y}_{1}^{*}(f))-h({Z}^{*}_{D+1}(f))\\&+ h(H(f) X_S^{*}(f)+{Z}_{D+1}^{*}(f)) - h({H}_{1,S}{X}_{S}^{*}(f)+{Z}^{*}_{1}(f))\Big)
\end{align*}
where, in the final inequality, the asterisk superscript indicates that the entropy is evaluated using a Gaussian distribution where the variance of $X_i(f)$ is $P_i^{[f]}$. In the inequality, the first term follows from the the chain rule, the facts that conditioning reduces entropy, and that the circularly symmetric Gaussian random variables maximize differential entropy under a power constraint. The second term follows simply from the distribution of the additive noise. The final two terms are bounded using the fact that $H_{1,S}(f)X_S(f)^T+Z_{1}(f)^T$ is a degraded version of $H(f)X_S(f)^T+Z_{D+1}(f)^T$ for all $f=1,2,\ldots,F$ and Lemma 2 in \cite{Cadambe_Jafar_Noisy}.
Thus, we can write 
\begin{align*}&C^{MAC-Z-BC}(F,\overline{P},\overline{\mathbf{H}}) \leq \sum_{f=1}^{F}\Bigg( \log\left(1+\sum_{j=1}^{S}|H_{1,j}(f)|^2P_j^{[f]}\right)\\&~~~~~~~~~~~~~~~~~~~~~~~~~~~~~~~+\log\left(\frac{1+|H(f)|^{2}P^{[f]}_S}{1+|H_{1,S}(f)|^2P^{[f]}_S}\right)\Bigg)\end{align*}
This completes the proof.
\section{Proof of Property 1}
\label{app:property}
Consider any network $\mathcal{N}$ characterized by $S,D,\mathcal{E},\mathcal{M}$. We split the proof into various cases.
\subsection*{Case 1 : $\mathcal{M} \neq \mathcal{E}$} In this case, $\mathcal{E}-\mathcal{M}$ is non-empty; without loss of generality, let $(1,1) \in \mathcal{E}-\mathcal{M}$. Also, based on the system model $\exists i_0,j_0$ such that $(1,i_0) \in \mathcal{M}$ and $(j_0,1) \in \mathcal{M}$. Now, the edges $\{(1,1),(1,i_0),(j_0,1)\}$ form a $Z$-interference network. Therefore, the $Z$-interference network is a sub-network of $\mathcal{N}$ if $\mathcal{M} \neq \mathcal{E}$.
\subsection*{Case 2 : $\mathcal{M} = \mathcal{E}$, and $\exists i_0 \neq j_0, \deg(T_{i_0}) > 1 < \deg(T_{j_0}) $}
Let $\mathcal{D}_{i} = \{j:(j,i) \in \mathcal{E}, j\in\{1,2,\ldots,D\}$ represent the set of destinations which are connected directly to source node $i$. Note that $|\mathcal{D}_{i_0}| > 1 < |\mathcal{D}_{j_0}|$. We divide this case into three sub-cases as follows.
\subsubsection{Sub-case (a) : $|\mathcal{D}_{i_0} \cap \mathcal{D}_{j_0}| \geq 2$}
In this case, let $i_1 \neq i_2, i_1,i_2 \in \mathcal{D}_{i_0} \cap \mathcal{D}_{j_0}$. Then, the edges $\{i_1,j_1\} \times \{i_0,j_0\} \subset \mathcal{E}=\mathcal{M}$ form a $2$-user $X$ network as a sub-network of $\mathcal{N}$. 
\subsubsection{Sub-case (b) - $|\mathcal{D}_{i_0} \cap \mathcal{D}_{j_0}| = 1$}
Here, let $k_1 \in \mathcal{D}_{i_0} \cap \mathcal{D}_{j_0}$. Also, let $i_1 \neq k_1 \neq  j_1$ and $i_1 \in \mathcal{D}_{i_0}, j_1 \in \mathcal{D}_{j_0}$. Then, it can be easily noted that the edges $ \{(i_1, i_0), (k_1,i_0),(k_1,j_0), (j_1,j_0) \} \subset \mathcal{E}$ form a $\Sigma$ network as a sub-network of $\mathcal{N}$. 
\subsubsection{Sub-case (d) - $\mathcal{D}_{i_0} \cap \mathcal{D}_{j_0} = \phi$} Let $i_1 \in \mathcal{D}_{i_0}$ and $j_1 \in \mathcal{D}_{j_0}$. Note that we consider connected network graphs in our model. Let $E_1,E_2,\ldots,E_N$ be a sequence of edges denoting a path, devoid of cycles, between nodes $i_1$ and $j_1$, where $E_n \in \mathcal{N}, n=1,2,\ldots,N.$ First, note that since the path is between two receivers (in our bipartite network graph), $N$ has to be even.  Now, note that if $N > 3$, this automatically means that the edges $E_1,E_2,E_3,E_4$ form a $\Sigma$ network. Therefore, we only need to consider the case where $N=2$. If $N=2$, let $E_1 = (i_1,k_{0})$ and $E_2=(j_1,k_0)$. Notice that we are considering the case where $\mathcal{D}_{i_0} \cap \mathcal{D}_{j_0} = \phi$  which means that $i_0 \neq k_0 \neq j_0.$ Then, the edges $\{(i_1,i_0),(i_1,k_0),(j_1,k_0),(j_1,j_0)\} \subset \mathcal{E}$ form a $\lSigma$ network as a sub-network of $\mathcal{N}$. 
\subsection*{Case 3 : $\mathcal{M} = \mathcal{E}$, and $\exists i_0 \neq j_0, \deg(R_{i_0}) > 1 < \deg(R_{j_0}) $}
The proof for this case is similar to the proof for case $2$.

This completes the proof.


\bibliographystyle{ieeetr}
\bibliography{Thesis}

\begin{thebibliography}{1}

\bibitem{Cadambe_Jafar_inseparable}
V.~Cadambe and S.~Jafar, ``Parallel {G}aussian interference channels are not
  always separable,'' {\em IEEE Transactions on Information Theory}, vol.~55,
  pp.~3983--3990, Sept. 2009.

\bibitem{Sankar_inseparable}
L.~Sankar., X.~Shang, E.~Erkip, and H.~Poor, ``Ergodic two-user interference
  channels: Is separability optimal?,'' in {\em Communication, Control, and
  Computing, 2008 46th Annual Allerton Conference on}, pp.~723--729, Sept.
  2008.

\bibitem{Nazer_Gastpar_Jafar_Vishwanath}
B.~Nazer, M.~Gastpar, S.~A. Jafar, and S.~Vishwanath, ``Ergodic interference
  alignment,'' in {\em Proceedings of IEEE International Symposium on
  Information Theory}, June 2009.

\bibitem{Cadambe_Jafar_maczbc}
V.~Cadambe and S.~Jafar, ``Sum-capacity and unique separability of the
  {MAC-Z-BC} network,'' 2009.
\newblock Preprint available on authors' website.

\bibitem{Jafar_Shamai}
S.~Jafar and S.~Shamai, ``Degrees of freedom region for the {MIMO} {X}
  channel,'' {\em IEEE Trans. on Information Theory}, vol.~54, pp.~151--170,
  Jan. 2008.

\bibitem{Chong_Motani_Garg_Z}
H.-F. Chong, M.~Motani, and H.~K. Garg, ``Capacity theorems for the "z"
  channel,'' {\em IEEE Transactions on Information Theory}, vol.~53, no.~4,
  pp.~1348--1365, 2007.

\bibitem{Cadambe_Jafar_X}
V.~Cadambe and S.~Jafar, ``Interference alignment and the degrees of freedom of
  wireless {X} networks,'' {\em IEEE Trans. on Information Theory,}, vol.~55,
  pp.~3893--3908, Sept. 2009.

\bibitem{Cadambe_Jafar_Noisy}
V.~R. Cadambe and S.~A. Jafar, ``Interference alignment and a noisy
  interference regime for many-to-one interference channels,'' {\em arxiv.org},
  vol.~abs/0912.3029, 2009.
\newblock http://arxiv.org/abs/0912.3029.

\end{thebibliography}
}
\end{document}